\documentclass[longauth]{aa}
\usepackage[varg]{txfonts}
\bibpunct{(}{)}{;}{a}{}{,} % to follow the A&A style
\usepackage{graphicx}	% Including figure files
\usepackage{amsmath}	% Advanced maths commands
\usepackage{natbib}
\usepackage{longtable}
\usepackage{rotating}
\usepackage{adjustbox}
\usepackage{lmodern}
\def\euclid{{\it Euclid }}
\def\roman{{\it Roman }}
\def\euclidnsp{{\it Euclid}}
\def\romannsp{{\it Roman}}
\def\repository{\protect\url{https://github.com/euclid-egbs/HistoricList}}

\begin{document}

\title{Historic microlensing events in the \euclid Galactic Bulge Survey}

\author{V. Bozza\inst{\ref{UNISA},\ref{INFN}}, L. Salmeri\inst{\ref{UNISA},\ref{OACN},\ref{UNINA}}, P. Rota \inst{\ref{UNISA},\ref{INFN}}, E. Bachelet \inst{\ref{Besancon}}, J.-P. Beaulieu\inst{\ref{Greenhill},\ref{IAP}}, A.A. Cole\inst{\ref{Greenhill}}, J.C. Cuillandre \inst{\ref{Saclay}}, E. Kerins\inst{\ref{Manchester}}, I. McDonald\inst{\ref{Manchester}}, P. Mr{\'o}z\inst{\ref{Warsaw}},
M. Penny \inst{\ref{Louisiana}}, C. Ranc\inst{\ref{IAP}}, N. Rektsini\inst{\ref{IAP}}, E. Thygesen\inst{\ref{Greenhill}}, H. Verma\inst{\ref{Louisiana}}, \\ The OGLE collaboration \\ A. Udalski\inst{\ref{Warsaw}},
R. Poleski\inst{\ref{Warsaw}},
J. Skowron\inst{\ref{Warsaw}},
M. K. Szyma{\'n}ski\inst{\ref{Warsaw}},
I. Soszy{\'n}ski\inst{\ref{Warsaw}},
P. Pietrukowicz\inst{\ref{Warsaw}},
S. Koz{\l}owski\inst{\ref{Warsaw}},
K. Ulaczyk\inst{\ref{Coventry}},
K.A. Rybicki\inst{\ref{Warsaw}},
P.  Iwanek\inst{\ref{Warsaw}},
M. Wrona\inst{\ref{Warsaw},\ref{Villanova}},
M. Gromadzki\inst{\ref{Warsaw}},
M.J. Mr{\'o}z\inst{\ref{Warsaw}}, 
\\The MOA collaboration \\ F. Abe\inst{\ref{Nagoya}},
D.P. Bennett\inst{\ref{Goddard},\ref{Maryland}},
A. Bhattacharya\inst{\ref{Goddard},\ref{Maryland}},
I.A. Bond\inst{\ref{Massey}},
R. Hamada\inst{\ref{Osaka}},
Y. Hirao\inst{\ref{Tokyo}},
A. Idei \inst{\ref{Osaka}},
S. Ishitani Silva\inst{\ref{Goddard},\ref{Maryland}},
S. Miyazaki\inst{\ref{Kanagawa}},
Y. Muraki\inst{\ref{Nagoya}},
T. Nagai\inst{\ref{Osaka}},
K. Nunota\inst{\ref{Osaka}},
G. Olmschenk\inst{\ref{Goddard}},
N.J. Rattenbury\inst{\ref{Auckland}},
Y.K. Satoh\inst{\ref{Kanagawa}},
T. Sumi\inst{\ref{Osaka}},
D. Suzuki\inst{\ref{Osaka}},
T. Tamaoki\inst{\ref{Osaka}},
S.K. Terry\inst{\ref{Goddard},\ref{Maryland}},
P.J. Tristram\inst{\ref{Canterbury}},
A. Vandorou\inst{\ref{Goddard},\ref{Maryland}},
H. Yama\inst{\ref{Osaka}}
}

\offprints{V. Bozza, \email{valboz@sa.infn.it}}

\institute{Dipartimento di Fisica "E.R. Caianiello", Universit\`{a} di Salerno, Via Giovanni Paolo II 132, Fisciano, I-84084, Italy \label{UNISA} 
\and Istituto Nazionale di Fisica Nucleare, Sezione di Napoli, Via Cintia, Napoli, I-80126, Italy \label{INFN}
\and Istituto Nazionale di Astrofisica - Osservatorio Astronomico di Capodimonte, Salita Moiariello 16, Napoli, Italy
\label{OACN}
\and Universit\`{a} degli studi di Napoli Federico II, Dipartimento di Fisica "Ettore Pancini", via Cinthia 21, Edificio 6, Napoli, Italy
\label{UNINA}
\and Universit\'e Marie et Louis Pasteur, CNRS, Institut UTINAM UMR 6213, Besan\c{c}on, France \label{Besancon}
\and Greenhill Observatory \& School of Natural Sciences, University of Tasmania, P.O. Box 807, Sandy Bay, TAS 7006 Australia \label{Greenhill}
\and Sorbonne Universit\'e, CNRS 
UMR 7095, Institut d'Astrophysique de Paris, 98 bis bd Arago,
75014 Paris, France \label{IAP}
\and Universit\'e Paris-Saclay, Universit\'e Paris Cit\'e, CEA, CNRS, AIM, 91191, Gif-sur-Yvette, France \label{Saclay}
\and Jodrell Bank Centre for Astrophysics, University of Manchester, Oxford Road, Manchester, M13 9PL, UK \label{Manchester}
\and Astronomical Observatory, University of Warsaw, Al.~Ujazdowskie~4, 00-478~Warszawa, Poland \label{Warsaw}
\and Department of Physics and Astronomy, Louisiana State University, Baton Rouge, LA 70803 USA \label{Louisiana}
\and Department of Physics, University of Warwick, Gibbet Hill Road, Coventry, CV4~7AL,~UK \label{Coventry}
\and Villanova University, Department of Astrophysics and Planetary Sciences, 800 Lancaster Ave., Villanova, PA 19085, USA \label{Villanova}
\and Institute for Space-Earth Environmental Research, Nagoya University, Nagoya 464-8601, Japan\label{Nagoya}
\and Code 667, NASA Goddard Space Flight Center, Greenbelt, MD 20771, USA\label{Goddard}
\and Department of Astronomy, University of Maryland, College Park, MD 20742, USA\label{Maryland}
\and School of Mathematical and Computational Sciences, Massey University, Auckland 0745, New Zealand \label{Massey}
\and Department of Earth and Space Science, Graduate School of Science, Osaka University, Toyonaka, Osaka 560-0043, Japan \label{Osaka}
\and Institute of Astronomy, Graduate School of Science, The University of Tokyo, 2-21-1 Osawa, Mitaka, Tokyo 181-0015, Japan \label{Tokyo}
\and Institute of Space and Astronautical Science, Japan Aerospace Exploration Agency, 3-1-1 Yoshinodai, Chuo, Sagamihara, Kanagawa 252-5210, Japan \label{Kanagawa}
\and Department of Physics, University of Auckland, Private Bag 92019, Auckland, New Zealand \label{Auckland}
\and University of Canterbury Mt.¥ John Observatory, P.O. Box 56, Lake Tekapo 8770, New Zealand \label{Canterbury}
}

\date{Received/ Accepted}

\abstract {
Microlensing campaigns have a long history of observations covering the Galactic bulge, where thousands of detections have been obtained, including many exoplanetary systems. The \euclid Galactic Bulge Survey represents a unique opportunity to revisit a large number of past events and attempt the lens-source resolution of known events falling in the covered area.}{As the analysis of individual events requires non-negligible efforts, it is important to establish priorities among all possible targets, identifying those candidates with the higher chance for a successful resolution of the lens from the source and with the highest scientific interest. }{Drawing from the databases of the three main microlensing surveys (OGLE, MOA and KMTNet), we compile the complete catalog of past microlensing events in the \euclid survey footprint up to year 2023, containing 7801 entries. By re-modeling all events and cross-checking with Galactic models, we estimate the relative lens-source proper motions for all events. }{
Taking into account all uncertainties, for each microlensing event we are able to estimate the probability that the lens is separated from the source by more than a given angular distance threshold. Hence, we rank all events by their resolution probability, providing additional useful information that will guide future analyses on the most promising candidates. A particular attention is dedicated to the 51 known planetary microlensing events.
}{}

\titlerunning{Historic microlensing events in the \euclid Survey}
\authorrunning{V.Bozza et al.}
\maketitle 

\section{Introduction}\label{Sec intro}

Gravitational microlensing refers to the temporary increase in the flux of a background source due to the deflection of its light by the gravitational field of an object passing across the line of sight \citep{Pazsynski1986}. Following Paczynski's intuition and description of the basic characteristics of the microlensing phenomenon, several astronomers formed collaborations supporting seasonal observational campaigns aimed at the detection of such transient phenomena \citep{Alcock1993,Udalski1993}. Crowded stellar fields such as the Galactic Bulge offer the highest chance of the alignment needed for microlensing to occur, since stars in the disk or even in the bulge itself may serve as gravitational lenses to background stars \citep{Udalski1994,Alcock2000,Afonso2003,Sumi2003,Sumi2013,Mroz2019,Nunota2025}. By the photometric measurement of the flux variation throughout the microlensing event, it is generally possible to obtain information about the lens system, regardless of the light from the lens itself, which may even be completely dark. For this reason, microlensing has been first employed to test the possibility that the dark matter was made of compact objects \citep{Alcock2000b,Tisserand2007,Wyrzykowski2011,Niikura2019,Mroz2024Nat,Mroz2024}, then to study stellar populations in the disk, in the bulge and nearby galaxies, and mostly for the detection of extrasolar planets orbiting the lens stars \citep{MaoPazynski1991,Bond2004,Gaudi2012,Tsapras2018,MrozPoleski2023}. More recently, a microlensing event has led to the discovery of the first official isolated black hole wandering across our Galaxy \citep{Sahu2022,Lam2022}.

The duration of a microlensing event is set by the so-called Einstein time $t_E$ (Eq. \ref{tE}), which is the basic observable parameter depending on the mass, distance and relative proper motion of lens and source. Such parameter alone is insufficient to clarify the real nature of an individual lens, since e.g. we may have a fast massive lens or a slow light lens leading to the same Einstein time. Without additional information, microlensing can only be employed in statistical studies on the whole population of lenses. For example, an excess in the number of short events could be due to a population of free-floating planets \citep{Mroz2017,Sumi2023}.

There are several ways to break this degeneracy for an individual event: the measurement of the annual parallax (Eq. \ref{piE}) is possible for long enough events \citep{Gould1992}, the finite size of the source star (Eq. \ref{rho}) intervenes if the impact parameter between the lens trajectory and the line of sight is small enough \citep{WittMao1994}, observations from a satellite may fix the relative parallax between lens and source \citep{Refsdal1966}.

In addition to these methods, we may attempt direct lens detection if we wait long enough to let the lens move away from the glare of the source \citep{Bennett2006,Beaulieu2018}. It is clear that such detection would immediately return the relative lens-source proper motion. In addition, the lens flux can be used as an additional constraint on the lens mass and distance through the use of stellar models. This technique has been successful in a dozen cases, for which Adaptive Optics (AO) at 10 m-class telescopes has been required or observations from the \emph{Hubble Space Telescope} (\emph{HST}). In several cases, such observations have been able to discriminate between different alternatives and drive microlensing models toward the correct physical description of the event \citep{Bennett2016,Vandorou2020,Bennett2024,Terry2024}. In all cases, the lens detection has dramatically shrunk the error bars in the mass and distance measurements of the lens \citep{Batista2014,Bennett2015,Bhattacharya2018,Bennett2020,Blackman2021,Bhattacharya2023,Vandorou2025b}.

Indeed, the resolution of the lens from the source is a key element in the strategy devised for the future Galactic Bulge Time Domain Survey (GBTDS) to be performed by the \roman space telescope\footnote{\protect\url{https://roman.gsfc.nasa.gov/}} starting from 2027 \citep{Bhattacharya2019}. In fact, by observing the same fields after a few years, \roman will likely detect the lenses responsible for most of the microlensing events found in the first season and vice versa. For such events, the uncertainty in the masses of the discovered exoplanets will be significantly reduced, contributing to meeting the science requirements set by the mission goals \citep{Terry2026}.

In the middle between ground-based surveys and the space-based survey by \roman comes the ESA \euclid mission\footnote{\protect\url{https://www.esa.int/Science_Exploration/Space_Science/Euclid}}, which shares many similarities with \roman and is an ideal precursor to the NASA mission \citep{EuclidOverview}. While engaged with its Wide and Deep surveys, devoted to discover the nature of Dark Energy, \euclid has taken a snapshot of the same fields that will be visited by \romannsp. With at least two years span between the \euclid images and the first \roman observations and a similar Point-Spread-Function (PSF) size, \euclid will further increase the chances of lens detections for future microlensing events \citep{Bachelet2022}.

However, the same idea can be implemented looking backward at past microlensing events discovered by ground-based surveys over the past 20 years. For many of the older microlensing events, enough time has passed to see the lens well-separated from the source in \euclid images. We thus have an invaluable opportunity to revisit interesting events, such as planetary or binary lenses, validate the past models and reduce the uncertainties of all mass measurements. Besides the published individual events, it is possible to undertake a much broader statistical study embracing all microlensing events for which we can detect the lens. Population studies can receive a validation by the real flux we see from the lens.  Even the non-detection of the lens can be of great value, since it may be used to prove that the lens is stellar remnant \citep{Blackman2021}. For short events that may be explained by a fast star or a slow free-floating planet, the second scenario would be favored in case of non-detection. 

For all these reasons, we should not miss the opportunity to use \euclid images to boost the overall microlensing collection to a higher level
of astrophysical information. As a start-up of this ambitious project, we should first identify all past microlensing events falling in \euclid fields and prioritize those that look most promising for the lens detection, trying to re-organize the existing information in such a way that is readily accessible to guide the incoming investigations. This is the goal of the present work. In section 2 we will first describe the \euclid Galactic Bulge Survey, we will then introduce the sources for past microlensing events in Section 3, presenting some overall statistics on microlensing events in \euclid fields. Section 4 will describe the methods we used to assess the probability for lens-source separation for all events. Section 5 brings the focus on published planets involved in this study, for which we have more information available. Finally, we draw the conclusions in Section 6.

\section{The \euclid Galactic Bulge Survey}

\begin{figure}
    \centering
    \includegraphics[width = 0.5\textwidth]{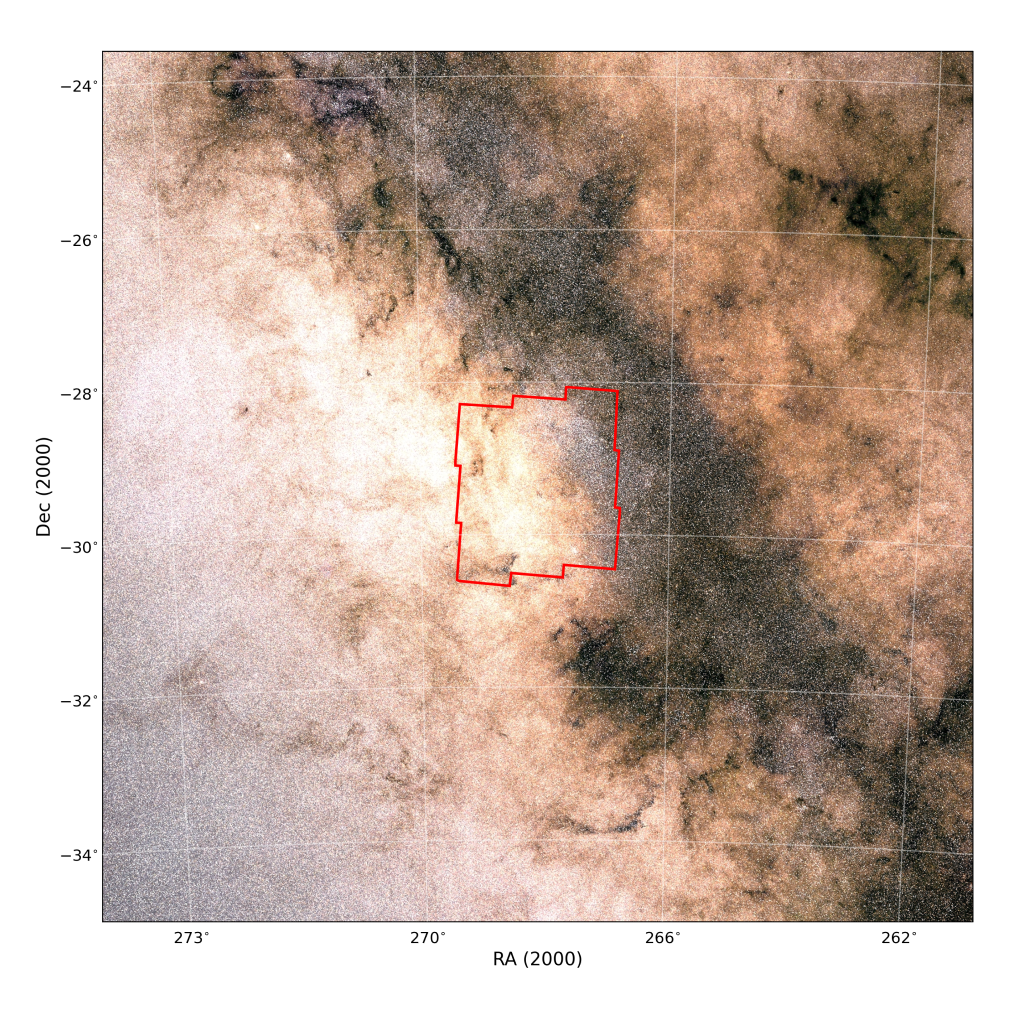}
    \caption{Outline of the nine Euclid Galactic Bulge Survey pointings (red) overlaid on a Gaia image of the Milky Way. The total outlined sky coverage is 4.8 deg$^2$.}
    \label{Fig mosaic}
\end{figure}

The \euclid mission was launched on 1st July 2023 and is now orbiting in the Lagrangian point L2, while conducting a survey of the deep sky aimed at an accurate reconstruction of the history of our Universe, clarifying the nature of dark matter and dark energy and the formation of large-scale structures \citep{EuclidOverview}. It is made of a Korsch 1.2m telescope with an effective field of view of 0.54 deg$^2$ \citep{Racca2016}. In order to keep the telescope and the payload at the temperature of 126K, the telescope is shielded against sunlight and the solar aspect angle must be maintained close to 90$^\circ$. Further constraints arise to avoid stray light in the optics and affect the survey strategy of the mission. Because of these pointing restrictions, the fields in the Galactic bulge of interest for microlensing can only be viewed in two windows of about 23 days around the equinoxes. 

\euclid is able to support the detection and characterization of extrasolar planets by microlensing in two ways: a coordinated time-domain survey of the Galactic Bulge simultaneous with \roman to be performed during the allowed pointing windows \citep{Penny2013,Bachelet2019}; a single snapshot taken well before the beginning of \roman observations \citep{Bachelet2022}. A simultaneous survey would allow immediate parallax determination for most microlensing events observed by the two missions, in particular short events that are amenable candidates for free-floating planets. Such survey could be proposed for an extended \euclid mission. A one-time survey of the Galactic Bulge has been performed between 23rd and 24th March 2025 and constitutes the \euclid Galactic Bulge Survey (EGBS), which is described in great detail by \citet{Beaulieu2026} and briefly introduced here.

\euclid has two wide-field imaging instruments: VIS and NISP, operating in visual and near-infrared bands respectively. NISP was not used in the course of EGBS, so here we briefly describe VIS only. VIS is made of 36 CCDs in a 6 × 6 array. Each CCD has 4132 × 4096 pixels in four quadrants, so the VIS images comprise $6.09\times10^8$ pixels in 144 quadrants \citep{Cropper2016}. The wavelength sensitivity of VIS is $0.53\mu m<\lambda< 0.92 \mu m$. The plate scale is $0.10''pixel^{-1}$ which leads to moderate undersampling of the PSF, whose typical width is of the order of 0.14'' in the images. Since the VIS band is relatively broad, we must also take into account the fact that the PSF is chromatic: redder sources will have a broader PSF. Careful modeling of the PSF is needed to achieve correct photometry and astrometry. For these reasons, the reference observing sequence (ROS) includes 4 consecutive exposures per-field dithered according to a characteristic S-shape pattern. Multiple dithered exposures of the same field allow for a much more accurate reconstruction of the PSF at subpixel level, paving the way to precision astrometry. 

%NISP operates in wavelengths between 0.95 and 2.0 $\mu m$ and would have been a precious complement to VIS in the near-infrared in many respects \citep{Hormuth2025}, but concerns about persistence issues after the imaging of the crowded fields in the bulge suggested to cautiously avoid exposing NISP in the course of the EGBS\footnote{We mention that preliminary studies by the NISP team demonstrated that an adequate recovery time of a few hours after the EGBS would be sufficient to remove any residual signal. These studies are encouraging in view of possible future \euclid surveys towards the bulge including NISP.}.

\begin{table}
\centering
\begin{tabular}{lll}
\hline
Field & RA($^\circ$) & Dec($^\circ$) \\
 \hline 
1 & 267.4250 & -30.0194 \\
2 & 267.4406 & -29.2595 \\
3 & 267.4560 & -28.4996 \\
4 & 268.2266 & -30.1299 \\
5 & 268.2374 & -29.3699 \\
6 & 268.2480 & -28.61 \\
7 & 269.0296 & -30.2366 \\
8 & 269.0355 & -29.4766 \\
9 & 269.0414 & -28.7166 \\
\hline 
\end{tabular}
\caption{Reference coordinates of the nine fields of the EGBS.}
\label{Tab coordinates}
\end{table}

The EGBS imaged a total of nine fields in Baade's window. The coordinates of the centers of the fields are reported in Table \ref{Tab coordinates}. The fields were rotated by $17^\circ$ from the north. Fig. \ref{Fig mosaic} shows the fields of the survey superposed on a Gaia image of the Milky Way \citep{Gaia2018}. Note that the Galactic center lies on the right, where the extinction becomes more and more severe. We note that some small blank spaces are left between the dithers and the interchip gaps of the VIS instrument. 

The exposure time of each dither was 400s. For those stars 
enjoying more dithers, the limiting magnitude reaches up to AB 25.

Finally, an additional field outside the bulge was pointed with the purpose of facilitating the reconstruction of the PSF at the time of the EGBS.  The data have been publicly released in June 2026 \citep{Beaulieu2026}.

\section{Ground microlensing surveys} \label{Sec observations}

Observational campaigns for the detection of microlensing events started back in 1992, with the first project, the Optical Gravitational Lens Experiment (OGLE) still ongoing \citep{OGLE1992}. After the first pioneering 10 years with only a few events detected per year, the detection of new microlensing events received a boost from automatic alert systems, which were implemented by OGLE, Microlensing Observations in Astrophysics (MOA) and later on by the Korean Microlensing Telescope Network (KMTNet). We will entirely rely on the databases from these three main surveys to compile our list of historic microlensing events. 

OGLE observes from the 1.3m Warsaw telescope located in Las Campanas Observatory in Chile. It has undergone several upgrades bringing it to the current phase IV, whose details are described by \citet{OGLEIV}. 
In our catalog we investigate all events detected by  OGLE-II, OGLE-III and OGLE-IV, ranging from year 1998 to 2023. Most of these events were alerted in real time thanks to the Early Warning System (EWS) \citep{EWS1994}, but we also include some additional events that were recovered offline \citep{Mroz2019,Wyrzykowski2015}. For the purpose of this paper, we have agreed to use the online photometry available at OGLE online archive\footnote{\protect\url{https://ogle.astrouw.edu.pl/ogle4/ews/ews.html}} and use re-reduced photometry only in exceptional cases, as explained in Sect. \ref{Sec re-modeling}. We note that OGLE photometry comes in standard calibrated Cousins I-band, which will represent our standard for our analysis.

MOA observes from the 1.8m MOA-II telescope located at the University of Canterbury’s Mount John Observatory in New Zealand \citep{MOA2001,Sumi2003}. It underwent a major upgrade in 2006, when it started high-cadence observations. We include all events from the MOA archive\footnote{\protect\url{https://moaprime.massey.ac.nz/moaarchive}} starting from year 2000. We also include some anomalous events that were found in offline searches\footnote{\protect\url{http://iral2.ess.sci.osaka-u.ac.jp/~moa/anomaly/bin/index.html}}. Similarly to OGLE, we use online photometry for MOA as well. This comes in a customized red band, encompassing standard R- and I-bands. Calibration of MOA data to standard filters requires additional steps that need to be tailored to specific events. In our analysis, therefore, we will incorporate MOA photometry only in the form of instrumental magnitudes.

KMTNet uses three identical 1.6 m telescopes at the Cerro Tololo Inter-American Observatory (CTIO) in Chile, the South African Astronomical Observatory (SAAO) in South Africa, and the Siding Spring Observatory (SSO) in Australia \citep{KMTNet} \citep{KMTNet}. As they started observations in 2015, their baseline to EGBS is reduced compared to the other surveys, but still the number of short events found by this survey encourages the use of these data as well. Online public photometry for KMTNet has been used\footnote{\protect\url{https://kmtnet.kasi.re.kr/ulens/}}. We have used their I-band photometry which has a sufficiently good calibration for our purposes. 

\section{Historic microlensing events in EGBS}

\begin{figure}
    \centering
    \includegraphics[width = 0.5\textwidth]{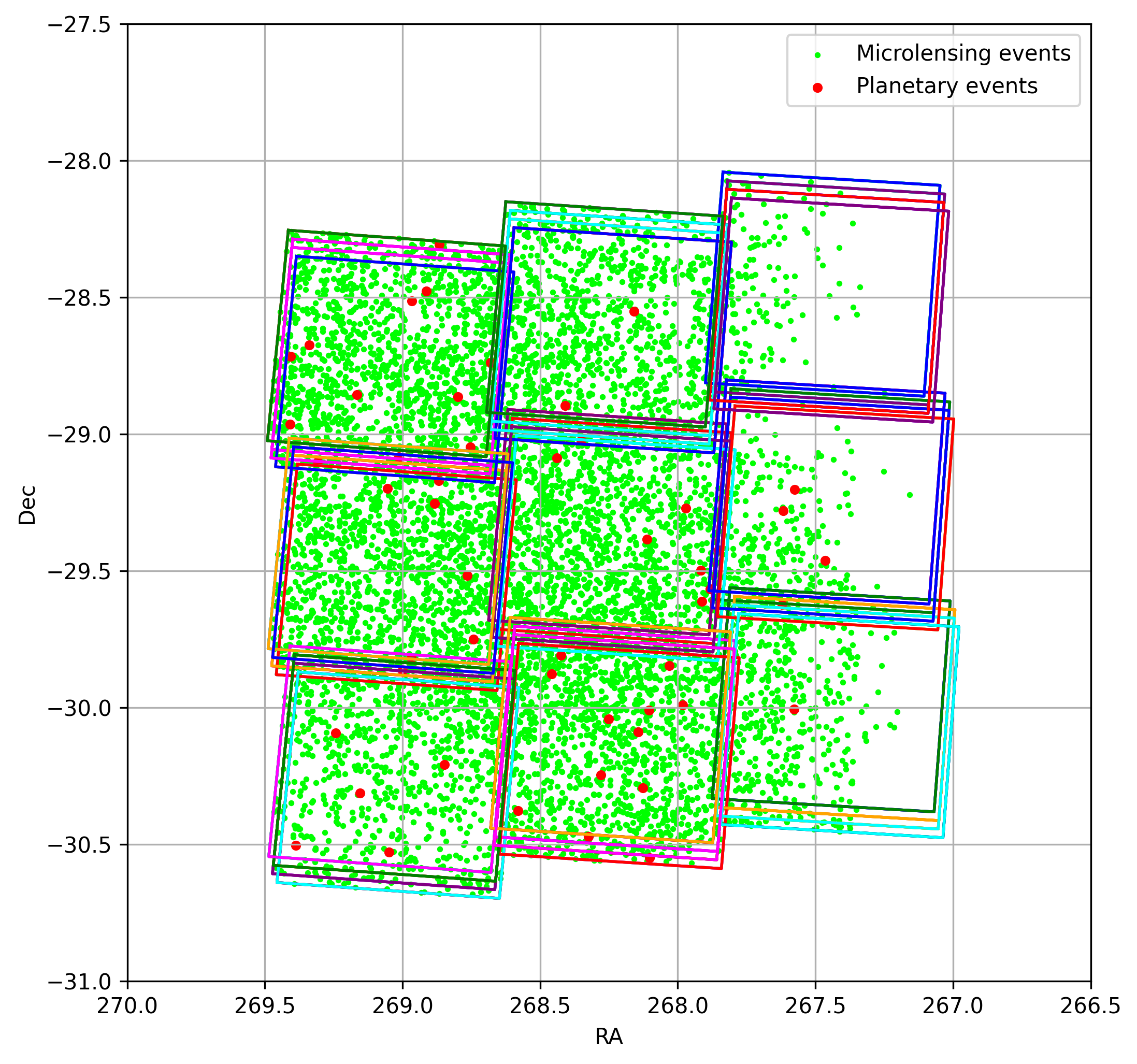}
    \caption{Fields observed in the EGBS. For each field four different dithers are visible. All historic microlensing events falling in the EGBS are shown in green. In particular, published planetary microlensing events are shown in red.}
    \label{Fig fields}
\end{figure}

The EGBS frames come with WCS astrometry for a precise reconstruction of the field coordinates. This allows us to easily check the presence of each of the past microlensing events discovered from the ground in the footprint of the EGBS. In the public repository \repository \; we have provided a script that returns the file names and the corresponding pixel coordinates in which an event in given astronomical coordinates appears. 

Using this tool, we have selected the historic events imaged by \euclid among all ground surveys, collecting {\bf 7801} independent events. They are shown in Fig. \ref{Fig fields} along with the EGBS fields. The decrease of events on the western side due to the increase of dust extinction as we approach the Galactic center is clearly visible. Within these events, we find 51 published planetary events, highlighted in red in the figure. These obviously constitute the targets of highest interest for the exoplanet community and will receive priority for the identification of the lenses. Among the most interesting events, we may mention OGLE-2006-BLG-109, the first multiplanetary event \citep{Gaudi2008,Bennett2010}, MOA-2011-BLG-293, with a possible lens detection by Keck that may lead to a candidate planet in the habitable zone awaiting further confirmation  \citep{Batista2014}, OGLE-2005-BLG-390, a system with an icy 5.5 Earth-masses planet \citep{Beaulieu2006}, the Earth-mass planet OGLE-2016-BLG-1195b \citep{Bond2017,Shvartzvald2017,Vandorou2025}, OGLE-2017-BLG-0173, a planet in a ``Hollywood'' { (large source size)} event \cite{Hwang2018}, the multiplanetary system OGLE-2018-BLG-1011 \citep{Han2019}, OGLE-2013-BLG-0341, a planet in a binary system \citep{Gould2014}, OGLE-2015-BLG-0966, OGLE-2018-BLG-0596 and OGLE-2017-BLG-0406, also observed by {\it Spitzer} \citep{Street2016,Jung2019,Hirao2020}, OGLE-2011-BLG-0462, the first isolated black hole discovered by microlensing \citep{Sahu2022,Lam2022}, the free-floating planet candidates KMT-2019-BLG-2073 \citep{Kim2021} and OGLE-2017-BLG-0560 \citep{Mroz2019}.

\begin{figure}
    \centering
    \includegraphics[width = 0.5\textwidth]{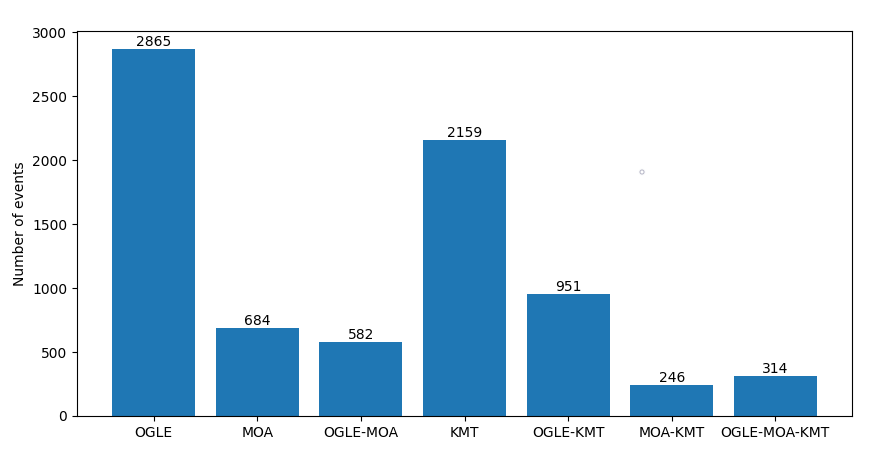}
    \caption{Distribution of events per survey. Some events have been observed by more than one survey.}
    \label{Fig surveys}
\end{figure}

\begin{figure}
    \centering
    \includegraphics[width = 0.5\textwidth]{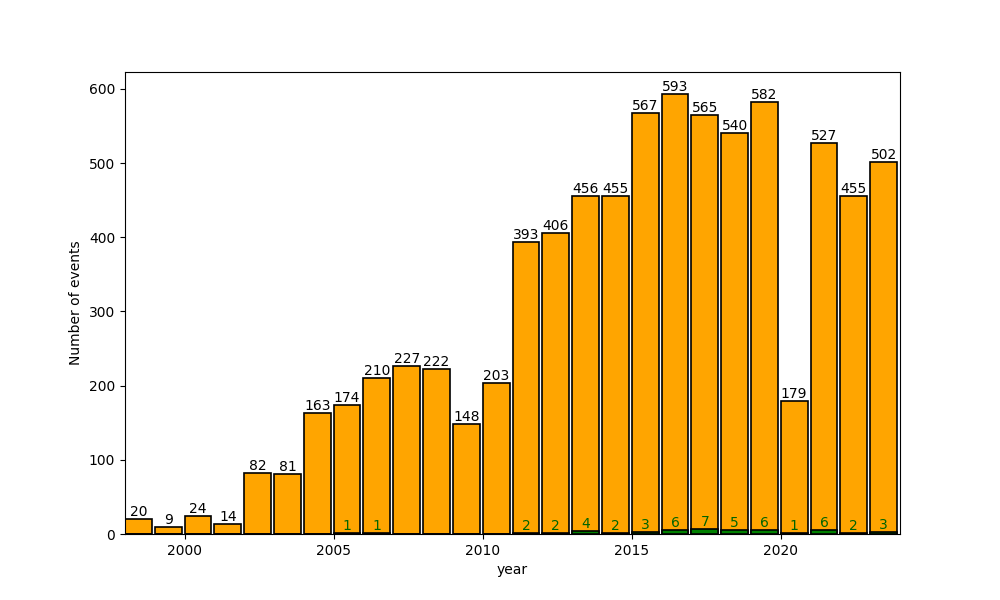}
    \caption{Distribution of events per year. In green the published planetary events.}
    \label{Fig years}
\end{figure}

Fig. \ref{Fig surveys} contains an overview of the events in EGBS divided by surveys. Some of the events have been observed by more than one survey. In this case, all the data are included in the re-analysis of Sect. \ref{Sec re-modeling}. Fig. \ref{Fig years} shows the distribution of events per year of discovery. We note that a first boost in the number of events came in 2011 with the onset of OGLE-IV and another boost in 2015 is due to the start of the KMTNet project. The interruption in 2020 due to the Covid pandemic is also evident.

For each of these events, the surveys provide synthetic information on their web pages. This includes coordinates, a finder chart, some parameters obtained by single-lens-single-source fitting: the baseline magnitude, the peak epoch, the Einstein time, the peak magnification or the impact parameter. Finally, some comments are sometimes posted about interesting or problematic events. For all events, online photometry is also available, which is the basis for our re-modeling.

\subsection{Re-modeling of events} \label{Sec re-modeling}

Our goal is to compile a ranking of all microlensing events in EGBS based on the probability that the lens can be resolved by \euclid observations. Therefore, we need a full homogeneous re-analysis of all microlensing events with the same modeling platform and with the same criteria, also including binary lens and/or binary source modeling. The main parameters we wish to extract are the Einstein time and, when measurable, the microlensing parallax, due to the orbital motion of the Earth. 

The Einstein time is the ratio between the Einstein angle $\theta_E$ and the relative lens-source proper motion $\mu_{geo}$ as seen from the Earth at the time of the microlensing event:
\begin{equation}
t_E \equiv \frac{\theta_E}{\mu_{geo}}. \label{tE}
\end{equation}

The Einstein angle fixes the lensing power of the event and is given by:
\begin{equation}
\theta_E \equiv \sqrt{\kappa M \pi_{rel}}, \label{thetaE}
\end{equation}
where $M$ is the mass of the lens, $\pi_{rel} = \pi_L-\pi_S$ is the difference of the geometric parallaxes of lens and source, and $\kappa = 4G/c^2\mathrm{au}$ combines the Newton constant, the speed of light and the astronomical unit.

The microlensing parallax 
\begin{equation}
\pi_E \equiv \frac{\pi_{rel}}{\theta_E} \label{piE}
\end{equation}
can be measured for long enough events for which the Earth orbital motion is relevant or by satellite observations.

Finally, the source physical size can affect the microlensing light curve sometimes. The parameter
\begin{equation}
\rho_* \equiv \frac{\theta_*}{\theta_E} \label{rho}
\end{equation}
can be useful if the angular size of the source $\theta_*$ can be estimated by stellar models. 

In general, in order to fully break the degeneracy between the proper motion $\mu_{geo}$, the lens mass $M$ and the distances to lens $D_L\equiv \mathrm{au}/\pi_L$ and source $D_S\equiv \mathrm{au}/\pi_S$, it is necessary to have a good measurement of $\pi_E$ and $\rho_*$, coupled with some hypothesis on the source through stellar models. It is clear that this is not possible for the vast majority of the microlensing events under examination in this investigation because finite-size effects are very rare, the microlensing parallax is less rare but not typically measured, in particular for short time-scale events, which are the most promising to be resolved by \euclidnsp. Finally, a color--magnitude diagram (CMD) could be used to investigate the nature of the microlensing source, so as to estimate $\theta_*$ or to infer the source distance, but not all events have available color information in the target field.

Therefore, we will just compile our ranking based on the simplest information we can extract from our models: $t_E$ and $\pi_E$ when possible, the baseline magnitude $I_{base}$ and the peak epoch $t_0$. The use of $\rho_*$ would require a CMD investigation for each event, which is clearly out of reach. The situation is different for published planetary events, for which the full investigation is already publicly available. We will dedicate a separate section (Section \ref{Sec planets}) to these events.

Having re-framed the scope of our re-modeling to $I_{base}$, $t_E$ and $\pi_E$, it is clear that the online photometry is generally sufficiently good to achieve such limited goal. In fact, the purpose of re-reduction by sophisticated pipelines is to improve the accuracy so as to reveal subtle features or anomalies in the light curve. While such work is of primary importance to detect and characterize planets correctly, it has very little impact on global parameters such as the time-scale of the event, which sets the Einstein time. The microlensing parallax may be affected by long-term systematics that could be present in online photometry, but given the lower priority of long time-scale events in our list, we accept the possibility that our estimate of parallax may be altered by systematics in some of these events. There is certainly no sufficient motivation to enterprise a titanic work of re-reduction of { 7801} events only to adjust a few probabilities that remain very low anyway.

The modeling platform chosen for the re-analysis of all microlensing events in EGBS is \texttt{RTModel}\footnote{\protect\url{https://github.com/valboz/RTModel}} \citep{RTModel}, which provides models in the single-lens-single-source, binary-source and binary-lens categories and proposes an assessment based on a comparison of the $\chi^2$ achieved by each model category. The initial conditions for fitting are taken from a grid search for single-lens models and from a template library for binary-lens models. The fitting algorithm is a Levenberg-Marquardt with a bumping mechanism preventing duplicates and enlarging the exploration of the parameter space. The code employed for the computation of microlensing magnification is \texttt{VBMicrolensing} \citep{VBM}.

{ To speed-up the analysis for each event, we only include the data collected in the year of the peak, the year before and the year after, re-binned down to a maximum of 4000 data points. Such re-binning, as performed by \texttt{RTModel} preserves light curve variations and only acts on constant sections of the light curve (typically along the baseline), so it has virtually no impact on the inferred $t_E$. Points deviating more than 10 $\sigma$ from the extrapolation from neighbor points are also removed as outliers. For fitting we have imposed a semi--gaussian prior disfavoring negative blending on OGLE and KMTNet allowing for arbitrary positive blending and a gaussian prior on parallax components with $\sigma_\pi = 0.3$. We have also restricted the search to $t_E<1yr$, coherently with the clipping of our data on 3 years. Finally, since binary sources tend to interpret remaining outliers as a secondary source with extremely small source size, we have also imposed $0.001<\rho_*<0.5$ in the search for binary source models.

We are aware that there is a fraction of events alerted as microlensing and included in our sample that could actually be generated by other non-microlensing phenomena, such as eruptive or cataclismic variables, flares, background Supernovae, and so on. Most of these transients can be distinguished by a classical asymmetry test, since they have a much steeper rise and a slower descent. Therefore, we label as suspected non-microlensing phenomena all those events for which all photometric datasets agree that the time span from the baseline to the peak in the rising part is less than half the time span between the peak and the baseline in the descent. We do not exclude these events from the analysis and still report all the results as for genuine microlensing events, but they are clearly marked in the final catalog as possible contaminants.}

For each event, we record the following information, useful for the estimate of the lens-source proper motion and the resolution probability:
\begin{itemize}
    \item best model category in the letter coding used by \texttt{RTModel} { with an additional * for suspected contaminants};
    \item $\chi^2/d.o.f.$
    \item Einstein time $t_E$;
    \item baseline magnitude $I_{base}$;
    \item microlensing parallax $\pi_E$;
\end{itemize}

\begin{figure}
    \centering
    \includegraphics[width = 0.5\textwidth]{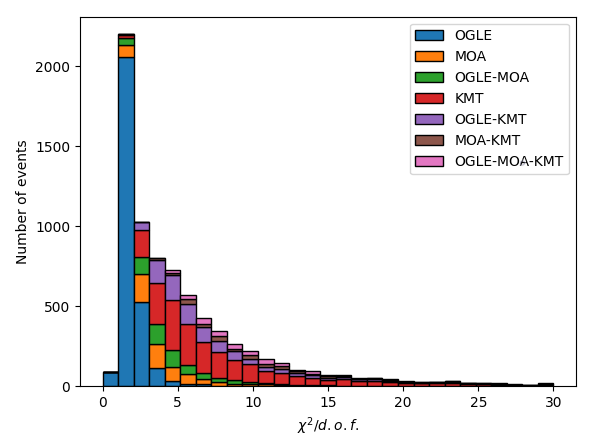}
    \caption{Distribution of events in terms of the $\chi^2/d.o.f.$ achieved in the fit.}
    \label{Fig chi2}
\end{figure}

{ The $\chi^2/d.o.f.$ can be considered as an assessment of the quality of the input online photometry and consequently the reliability of the model. A high value can be determined by the presence of intrinsic scatter in the data, outliers, by an inaccurate model or by the presence of a non-microlensing signal (periodic or eruptive variables). Fig. \ref{Fig chi2} shows the distribution of this indicator on the modeled events. We notice that all OGLE events are modeled with a $\chi^2/d.o.f.\sim 1$, indicating that our modeling is very efficient and robust on cleaner datasets. Online MOA photometry contains more outliers and is more noisy, but still all events are modeled with $\chi^2/d.o.f.< 10$. For KMTNet photometry, we find a long tail extending up to 30, due to an overall lower quality. Without a preliminary deep cleaning of these datasets, we cannot expect better values of the $\chi^2$, but still these data are precious because they contribute a substantial fraction of the total microlensing events and fill the gaps between other surveys.}

\begin{figure}
    \centering
    \includegraphics[width = 0.3\textwidth]{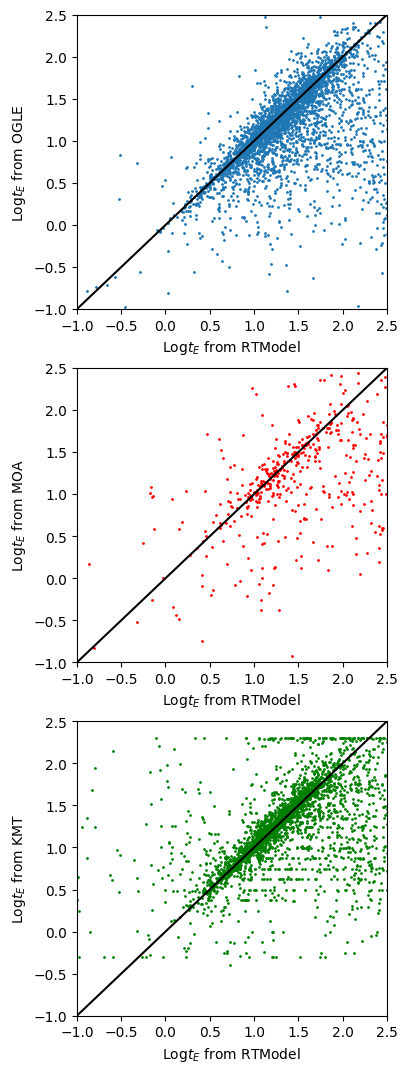}
    \caption{Comparison between the $t_E$ derived by our re-modeling with \texttt{RTModel} and the original $t_E$ reported in the online modeling by the surveys. The artificial horizontal lines in the KMT plot is due to the fact that only two significant digits have been reported for this parameter on the website.}
    \label{Fig tE comparison}
\end{figure}

{ It is interesting to see how much the original Einstein times reported by the surveys in their online analysis changes after our re-modeling. Fig. \ref{Fig tE comparison} shows a comparison between the Einstein time $t_E$ found by \texttt{RTModel} versus the Einstein time as found in the online analysis for each of the three surveys. For all three collaborations, we find a good agreement for most of the events, but there is a non-negligible fraction of events for which the Einstein times are very different. Looking closely at some of these, we find several different cases: 
\begin{itemize}
    \item Partially covered events (only the rise or only the descent) for which a good estimate of the Einstein time is impossible.
    \item Poorly sampled events, which are the majority. For these events the Einstein time is strongly degenerate with the blending and the impact parameter. The reported error bars are huge and actually make the old and new estimates compatible. In the OGLE plot, we may notice a tendency for \texttt{RTModel} to report a longer Einstein time compared to OGLE because OGLE gives priority to the results obtained with zero blending in such cases.
    \item Very noisy events, which have a naturally large uncertainty in the Einstein time.
    \item Problematic or non-microlensing events as tagged by our filter.
    \item Binary-lens or binary-source events or even single-lens-single-source with a significant parallax signal, for which the original estimates of $t_E$ obtained by the Paczynski light curve may be fairly distant from the true value after careful re-modeling.
    \item Events for which the use of multiple surveys change the estimate of the Einstein time.
\end{itemize} 

We note that the main purpose of our re-modeling is to address the last two classes of events, while the other classes will naturally come with large error bars tracking the impossibility to derive significant information on $t_E$ based on the historic data alone. 
}

\begin{figure}[t]
    \centering
    \includegraphics[width = 0.4\textwidth]{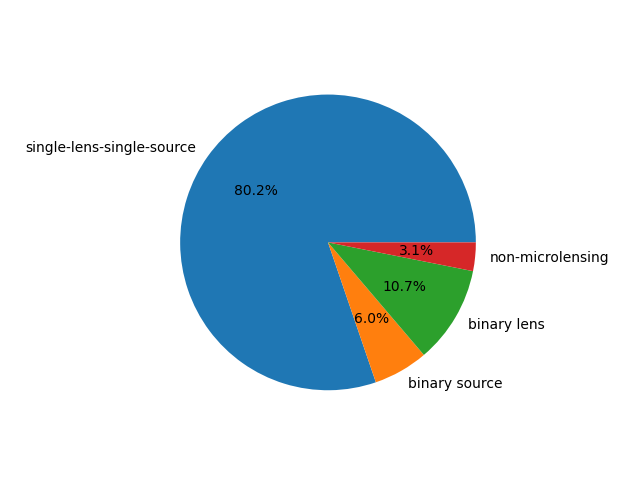}
    \caption{Incidence of different model categories as found by \texttt{RTModel}.}
    \label{Fig model categories}
\end{figure}

\begin{figure}[t]
    \centering
    \includegraphics[width = 0.4\textwidth]{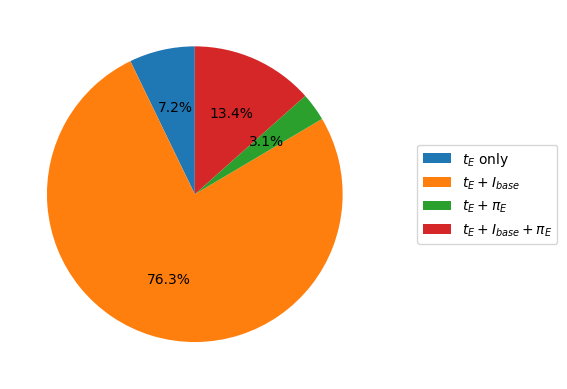}
    \caption{Events classified according to measured observables.}
    \label{Fig observables}
\end{figure}

The numbers of events falling in the three main categories (single-lens-single-source, binary-source, binary-lens) are reported in Fig. \ref{Fig model categories} { together with the events marked as non-microlensing}. As expected, the majority of events is made by single-lens-single-source, { with a $16.7\%$} classified as binary-lens or binary-source. Non-microlensing events amount to $3.1\%$ as found by our filter. Besides genuine binaries, it is certainly possible that scatter in the baseline can sometimes mimic an anomaly that is reported as a signature of a binary-lens or a binary-source. This is a consequence of the adoption of online photometry, which is naturally plagued by outliers or noise that would be removed by targeted photometry. \texttt{RTModel} obviously finds a decrease in $\chi^2$ by adding lenses and sources on problematic points of these light curves. { By excluding too small sources, we have dramatically reduced the impact of such spurious peaks obtaining a more robust classification. The residual presence of some misclassified events does not spoil the measurement of the Einstein time, which is fixed by the longer-scale genuine microlensing signal.}

We were not able to extract $I_{base}$ and $\pi_E$ for all microlensing events. As already anticipated, for MOA-only events we are unable to have a good enough calibration to estimate $I_{base}$. The microlensing parallax is measurable only for a minority of events. Fig. \ref{Fig observables} quantifies the numbers of events based on the measured parameters. { About 76\%} of events are assessed based on the measured $t_E$ and baseline magnitude $I_{base}$. Parallax is measured in { 16\% of events}.

\begin{figure}[t]
    \centering
    \includegraphics[width = 0.5\textwidth]{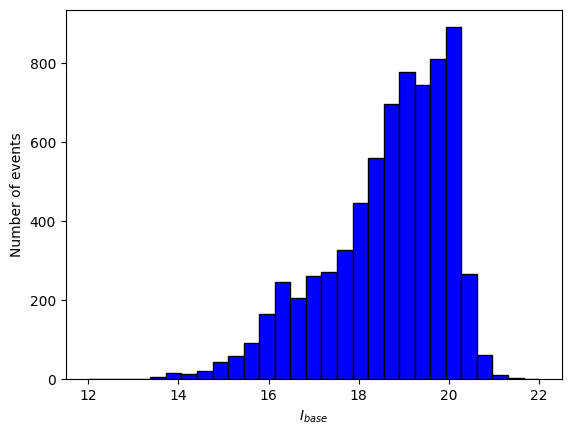}
    \caption{Distribution of events in terms of the baseline magnitude in I-band.}
    \label{Fig baseline}
\end{figure}

\begin{figure}[t]
    \centering
    \includegraphics[width = 0.45\textwidth]{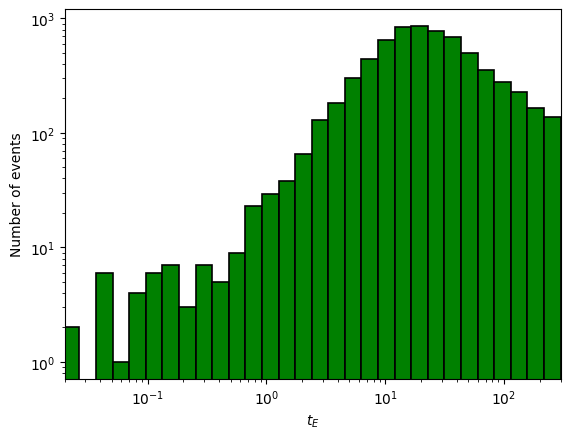}
    \caption{Distribution of events in the Einstein time $t_E$.}
    \label{Fig tE}
\end{figure}

\begin{figure*}[t]
    \centering
    \includegraphics[width = 0.8\textwidth]{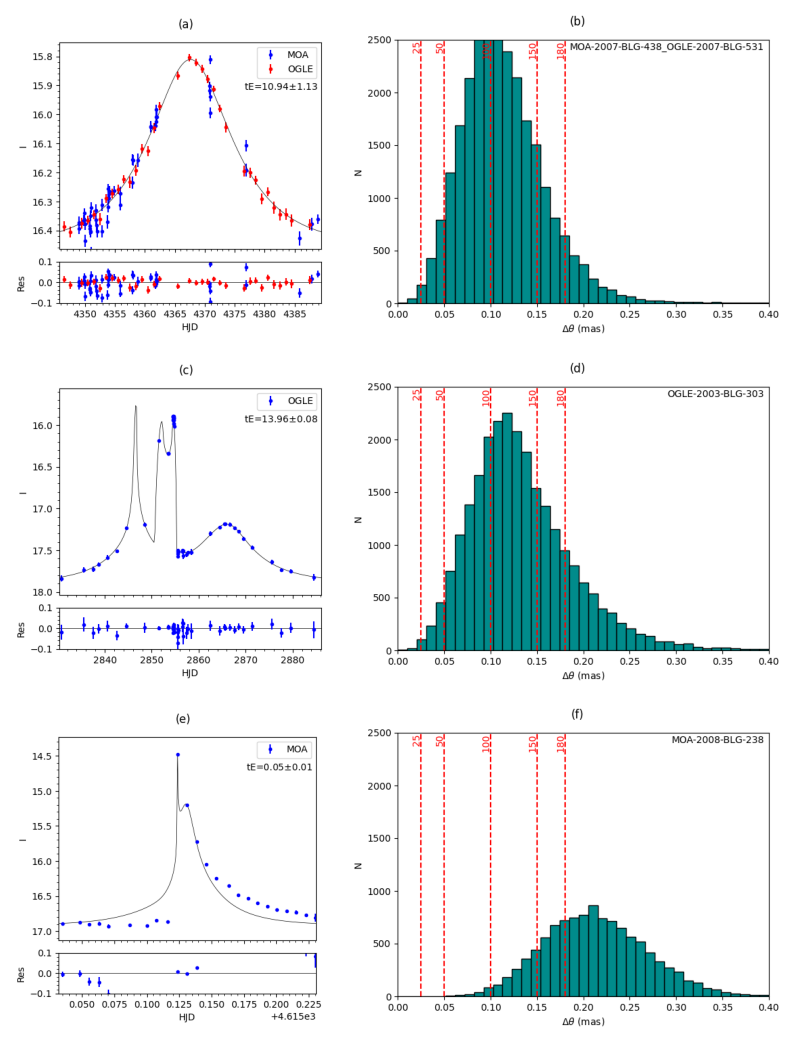}
    \caption{Three representative historic events. (a) Light curve for the single-lens-single-source event MOA-2007-BLG-438/OGLE-2007-BLG-531 and (b) the probability distribution for the lens-source separation for this event as calculated by \texttt{genulens} using the Einstein time obtained by \texttt{RTModel}. We have also marked the five thresholds used in the compilation of Table \ref{Tab Historic}. (c)-(d) refer to the binary lens OGLE-2003-BLG-303. (e)-(f) MOA-2008-BLG-238 is an example of a non-microlensing event in the historic sample, probably triggered by a flaring star.}
    \label{Fig events}
\end{figure*}

Fig. \ref{Fig baseline} shows the distribution of events in terms of the baseline magnitude $I_{base}$. This reflects the luminosity function in the bulge, with a decline for $I>20$ following the difficulty of performing accurate photometry from the ground for such faint sources.

Fig. \ref{Fig tE} contains the distribution of the events in the Einstein time $t_E$, whose Gaussian shape peaking at 20 days closely resembles similar distributions obtained by the surveys \citep{Sumi2023,Mroz2017}. Only the events with an uncertainty on $t_E$ lower than $t_E$ itself are displayed, { thus excluding 448 events. We also remind that we limited our search to Einstein times shorter than one year, so 394 longer events have been dropped from this histogram.} On the other side, we note that there are some very short events, which are associated with faster lens-source relative proper motion leading to higher chances for the resolution in \euclidnsp. These events may be also due to free-floating planets and thus deserve to be treated with the highest care.

\subsection{Galactic model and estimated proper motion}

After the re-modeling described in the previous sub-section, for each microlensing event in EGBS we now have robust estimates of the basic parameters, such as Einstein time, baseline magnitude and microlensing parallax. As said before, this is not sufficient to fully break the degeneracy between mass, distance and proper motion. In order to make a prediction on the probability of separating the lens from the source, we must rely on prior expectations from our knowledge of the Galactic stellar populations. Their space distribution and their kinematics provide guidance on the most likely ranges for the proper motion of each microlensing event. Crossing this prior expectation with the constraints coming from our measured parameters, we can derive posterior probability distributions for the proper motion that can be used to infer the expected lens-source separation for each microlensing event.

There are several Galactic models designed for such a Bayesian approach in microlensing analysis (see e.g. \citet{Robin2003,Han2003,Dominik2006,Yang2021}). A massive data analysis like the one we have to do for { 7801} microlensing events requires a fast, easy and reliable code that can be easily scripted. We have adopted the software \texttt{genulens}\footnote{\protect\url{https://github.com/nkoshimoto/genulens}} \citep{Zenodo2021}, based on the model by \citet{Koshimoto2021}, which incorporates stellar density and kinematic information from \emph{Gaia} \citep{Gaia2018}, OGLE \citep{Nataf2013,Mroz2019} and radial velocity surveys \citep{Clarke2019,Kunder2012AJ....143...57K} for a better characterization of the bulge.

For each microlensing event, we provide \texttt{genulens} with all available information that affects the proper motion distribution. Firstly, the Galactic coordinates determine the line of sight and the stellar populations potentially contributing as sources or lenses.  The Einstein time $t_E$ and the microlensing parallax $\pi_E$, when available provide basic constraints on the velocities, mass and distance of the lens through Eqs. (\ref{tE}-\ref{piE}). The baseline magnitude $I_{base}$ places an upper limit to the mass for both the source and the lens (if it is a star). In order to exploit this information, we must also estimate the extinction in the I-band. Since we cannot construct a CMD for all microlensing events, we rely on existing extinction maps. { In particular, we have chosen the OGLE extinction map archive\footnote{\protect\url{https://ogle.astrouw.edu.pl/cgi-ogle/getext.py}} \citep{Nataf2013}, which represents a standard for bulge observations.}

\subsection{Lens detection probabilities in historic events}

In the end, for each microlensing event, we get the distribution of the heliocentric proper motions compatible with the characteristics of our event. The posterior distribution for the heliocentric proper motion can be immediately translated to a distribution in the expected lens-source separation after multiplying the proper motion by the time baseline between the microlensing peak epoch $t_0$ and the EGBS time $HJD\simeq 2460758\pm 1$. At this point, we are in the position to estimate the probability that the lens-source separation exceeds our desired threshold. A few representative examples are shown in Fig. \ref{Fig events}. { We note that the distributions in the predicted separation $\Delta\theta$ are relatively broad in all cases. This reflects the shape of the distribution in $\mu_{hel}$ which always remains with a width $\Delta \mu_{hel} \sim \mu_{hel}$. Without a measurement of the Einstein angle $\theta_E$, the constraints from $t_E$ and (when available) $\pi_E$ are useful to set the order of magnitude of $\mu_{hel}$, but are unable to fix its value with a precision better than $50\%$. The width of the distribution is then determined by the Galactic model, which contains the information about the physical ranges for lens and source velocities and distributions in mass and distance.}

\begin{table*}
\centering
\setlength{\tabcolsep}{4pt}
\tiny 
\caption{An excerpt of the list of historic microlensing events in EGBS footprint. The full table is available at \repository. For each event we report the probabilities (percentages) that the lens-source separation exceeds the indicated threshold in mas. We also report the model inferred by \texttt{RTModel} (single-lens-single-source (PS), binary lens (LS), binary source (BS){ , a $*$ indicates a suspected non-microlensing event}), the Einstein time $t_E$ in days, the baseline magnitude, if available, the peak time $t_0$ in $HJD-2450000$, the extinction $A_I$ and the $\chi^2$/dof.}
\label{Tab Historic}
\begin{tabular}{lccccccccccc}
\hline\hline
Event & P(>180) & P(>140) & P(>100) & P(>50) & P(>25) & Model & $t_E$ & $I_{\rm baseline}$ & $t_0$ & $A_I$ & $\chi2/dof$ \\
\hline
MB-08-238 & 74.3 & 92.88 & 99.31 & 99.99 & 100.0 & PS* & $0.05\pm 0.00$ & NA & 4615.131 & 2.272 & 5.6 \\
MB-00-003 & 69.83 & 87.71 & 97.14 & 99.91 & 100.0 & PS & $3.04\pm 0.29$ & NA & 1692.29 & 1.878 & 1.35 \\
OB-99-20 & 68.0 & 86.39 & 96.64 & 99.88 & 100.0 & PS & $3.99\pm 1.15$ & 15.2 & 1317.073 & 1.904 & 8.52 \\
MB-00-007 & 67.98 & 86.96 & 96.88 & 99.92 & 100.0 & PS & $3.37\pm 0.17$ & NA & 1725.759 & 1.46 & 1.72 \\
MB-01-054 & 67.6 & 86.65 & 96.74 & 99.96 & 100.0 & BS & $0.01\pm 1.88$ & NA & 2219.537 & 1.295 & 4.88 \\
OB-99-34 & 62.02 & 82.96 & 95.52 & 99.84 & 100.0 & PS & $5.82\pm 0.13$ & 16.4 & 1369.641 & 2.274 & 3.45 \\
OB-05-100 & 61.59 & 83.5 & 96.1 & 99.91 & 100.0 & PS & $0.90\pm 0.41$ & 17.4 & 3453.15 & 2.536 & 2.35 \\
OB-00-52 & 61.59 & 82.91 & 95.55 & 99.83 & 100.0 & PS & $4.43\pm 1.53$ & 17.7 & 1752.971 & 1.521 & 2.16 \\
OB-99-41 & 59.53 & 81.09 & 95.14 & 99.83 & 100.0 & PS & $6.16\pm 0.31$ & 15.6 & 1397.784 & 1.424 & 2.92 \\
OB-98-17 & 56.15 & 78.58 & 93.6 & 99.77 & 99.99 & BS & $8.83\pm 1.61$ & 16.3 & 949.6464 & 1.681 & 2.33 \\
OB-98-35 & 55.63 & 78.28 & 93.69 & 99.73 & 99.99 & PS & $8.17\pm 0.23$ & 18.5 & 1059.537 & 2.385 & 2.21 \\
MB-09-450 & 53.69 & 82.88 & 97.27 & 99.99 & 100.0 & PS & $0.14\pm 0.02$ & NA & 5064.017 & 1.458 & 2.62 \\
MB-01-042 & 53.68 & 77.54 & 93.96 & 99.77 & 100.0 & PS & $5.30\pm 0.43$ & NA & 2159.675 & 1.766 & 3.53 \\
OB-98-18 & 52.58 & 75.41 & 92.37 & 99.68 & 99.99 & PS & $10.10\pm 1.29$ & 15.4 & 970.7382 & 1.378 & 1.32 \\
OB-02-070 & 52.5 & 76.71 & 93.11 & 99.72 & 99.99 & BS & $5.10\pm 0.81$ & 15.8 & 2387.621 & 3.318 & 2.14 \\
OB-02-328 & 52.32 & 76.76 & 93.52 & 99.74 & 99.99 & PS & $4.73\pm 0.34$ & 17.2 & 2517.619 & 2.899 & 1.84 \\
MB-02-002 & 51.38 & 76.14 & 93.19 & 99.72 & 99.99 & PS & $4.91\pm 2.45$ & NA & 2343.708 & 1.789 & 1.64 \\
OB-98-20 & 51.19 & 74.0 & 90.99 & 99.47 & 99.99 & PS & $11.66\pm 3.18$ & 16.8 & 967.2214 & 3.039 & 1.3 \\
OB-02-354/OB-02-361 & 50.4 & 75.92 & 93.33 & 99.76 & 99.99 & PS & $4.31\pm 0.32$ & 16.4 & 2544.73 & 2.331 & 6.93 \\
OB-06-146 & 50.29 & 77.39 & 94.48 & 99.9 & 100.0 & PS & $1.13\pm 0.06$ & 17.3 & 3837.851 & 1.823 & 1.61 \\
\hline
\end{tabular}
\end{table*}

Table \ref{Tab Historic} contains an excerpt of 20 events from the full list of { 7801} historic microlensing events available at \repository. For each event we report the modeling parameters together with the probabilities that the lens-source separation exceeds { five reference thresholds: 0.18'', corresponding to the diffraction limit on the red side of the VIS band; 0.14'', corresponding to the average FWHM of the PSF for the VIS instrument; 0.10'', the pixel angular scale in VIS; 0.050'', half-pixel scale; and 0.025'', a quarter pixel scale}. The real threshold for a successful detection of the lens as a separate object from the source depends on the relative flux ratio, the total flux, the color difference between source and lens, the crowding, the presence of the target on multiple dithers. For this reason, we have provided different thresholds that may apply to different cases. 

{ Some particular attention is deserved by the flux ratio between the lens and the source. It is well known that the ability to resolve and determine the separation degrades as the flux ratio becomes very unbalanced between the two objects. A guide formula for the achievable signal-to-noise was proposed by \citet{Bennett2007} and broadly used in the community. In principle, we may use the blending factor retrieved by our models and try to implement a detection threshold based on this formula. However, the EGBS was taken in the VIS instrument in a wide band, which has no direct correspondence with the historic survey bands. In order to transform the blendings in the surveys to the VIS filter a detailed color information on the lens and the source would be needed, but this is available only very sparsely. As anticipated in Sections \ref{Sec observations} and \ref{Sec re-modeling} we only use one band per survey. For these reasons, for each event we only provide the separation probability, which is the fundamental but not unique ingredient to achieve detection.}

\begin{figure}
    \centering
    \includegraphics[width = 0.5\textwidth]{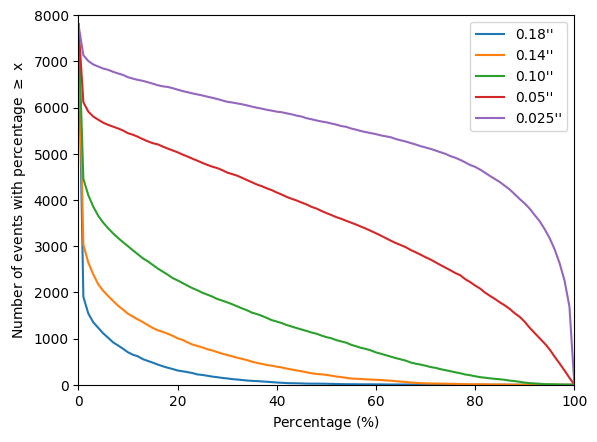}
    \caption{Number of events with separation probability higher than the given value in the abscissa. Each curve is drawn for a given reference detection threshold.}
    \label{Fig cumulant}
\end{figure}

Some interesting overall statistics can be drawn from this table. Fig. \ref{Fig cumulant} shows the number of events with { separation probability higher than the value in the abscissa assuming one of the five reference threshold}. All curves start from { 7801, since all events have a probability higher than 0\%, and terminate to zero at 100\%. For example, we can see that the number of events with probability higher than 50\% of having lens and source separated by more than 0.025'' is 5680. In other words, we expect 73\% of all microlensing events to be separated by more than 1/4 pixel. Similarly, 48\% of all events have a higher than 50\% probability of being separated by more than 0.050'', and still 13\% of events should be separated by more than 0.10'', corresponding to one full pixel}. Such values remind us that a generic historic microlensing event will still have lens and source unresolved in \euclidnsp. This also enforces the need of identifying the most promising microlensing events which should be considered in the first place in our analysis. Table \ref{Tab Historic} is sorted by decreasing resolution probability, with the most promising candidates on top of the list. Indeed, the top lines are dominated by short events that have been observed more than 20 years ago. { If these events are due to stellar lenses, they should be very likely resolved in EGBS. A non-detection of the lens in such short events would impose severe limits on the mass of a stellar candidate. For example, for MOA-2000-BLG-003, imposing $I_{lens}>25$, stellar lenses are limited below 0.3$M_\odot$ and confined in the bulge. Taken individually, this is not yet a conclusive evidence that this lens is a free-floating planet or a fast dark remnant, but placed in a statistical context it would greatly contribute to support the existence of a free-floating planet population \citep{Sumi2011,Mroz2017,Koshimoto2023}}.

\begin{figure}
    \centering
    \includegraphics[width = 0.5\textwidth]{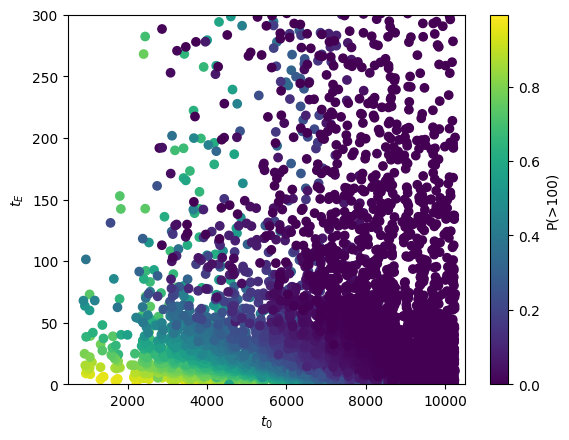}
    \caption{Probability of a lens-source separation larger than one \euclid pixel for historic events placed in the plane $(t_0,t_E)$.}
    \label{Fig t0tE}
\end{figure}

\begin{figure}
    \centering
    \includegraphics[width = 0.5\textwidth]{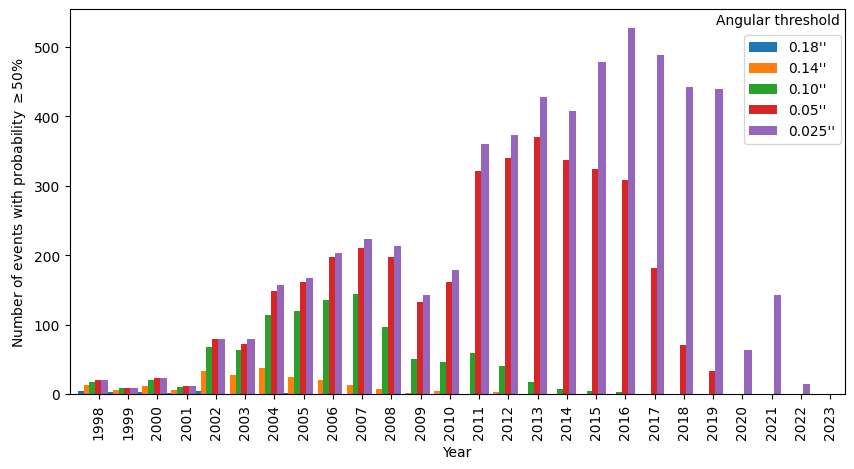}
    \caption{Number of events with detection probability higher than 50\% by year. { The  five colors refer to the different thresholds reported in the legend.}}
    \label{Fig nbyyear}
\end{figure}

{ 
The prediction for the lens-source separation basically depends on the time of the peak $t_0$ and the Einstein time $t_E$. Older and shorter events should have the largest separation. We can easily visualize this fact in Fig. \ref{Fig t0tE}, where the bottom-left corner enjoys the highest probabilities for separations above one \euclid pixel. }

Fig. \ref{Fig nbyyear} shows the distribution of the events with probability higher than 50\% for each reference threshold per year of detection. We see that higher thresholds are only accessible to older events, while lower thresholds still allow some more recent events to pass the selection.

{ Our results have some sensitivity to the specific choice of the Galactic model. We have adopted the default parametrization in \texttt{genulens}, which uses the form designed as $E+E_X$ for the Galactic bulge as described by \citet{Koshimoto2021}. We can compare our results with other parametrizations available in this software, such as form $G$ for the bulge \texttt{genulens}. We find that the differences in the detection probabilities stay well below $2\%$ in the majority of events. Only for shorter events ($t_E\lesssim 50d$), we observe a systematic increase in the detection probability with a median of $+4\%$ with the alternative bulge form. Similarly, we have also modified the low-mass regime of the Initial Mass Function, pushing the lower cut-off down to $10^{-6}M_\odot$. In this case, we only see a deviation for $t_E<1d$. For such short events, the detection probabilities decrease by $\sim 20\%$. Indeed, lower mass objects justify shorter times without the need of high proper motions. So, this trend can also be expected. We conclude that systematic uncertainties due to our limited knowledge of the Galaxy are of the order of a few percents, while the sensitivity of the predictions for short events to the low-mass end of the Initial Mass Function reiterates the importance of their investigation for a deep comprehension of the population of substellar objects and the unique opportunity offered by the EGBS.}

\section{Known microlensing planets} \label{Sec planets}

\begin{table*}
\centering
\caption{An excerpt of the table of planetary events falling in EGBS footprint. The full table is available online at \repository. For each event we report the expected lens-source separation; if more than one model exists, we report all possibilities. The method used in the reference paper to constrain the proper motion can be Finite-Source (FS), parallax (PX), High-Resolution imaging (HR), or a combination of these methods.}
\label{Tab planets}
\begin{tabular}{lccc} \\
\hline\hline
Event & $\delta\theta$ (arcseconds) & Method & Reference \\
\hline

OB050390 & $0.142 \pm 0.035$ & FS & \cite{Beaulieu2006}\\
MB140547-OB141760 & $0.133 \pm 0.001$ & HR & \cite{MB140547}\\
MB130260-OB130341\_M1 & $0.127 \pm 0.010$ & PX & \cite{Gould2014} \\
KB160212\_M2 & $0.119 \pm 0.039$ & FS& \cite{KB160212} \\
KB160212\_M3 & $0.098 \pm 0.030$ & FS & \\
KB160212\_M4 & $0.088 \pm 0.030$ & FS&  \\
KB160212\_M1 & $0.073 \pm 0.023$ & FS&  \\
OB151771\_M3 & $0.100 \pm 0.014$ & FS& \cite{OB0151771} \\
OB151771\_M2 & $0.089 \pm 0.017$ & FS& \\
OB151771\_M1 & $0.082 \pm 0.015$ & FS & \\
KB160372-OB161195-MB160350\_M1 & $0.097 \pm 0.004$ & HR& \cite{Vandorou2025} \\
KB160372-OB161195-MB160350\_M2 & $0.096 \pm 0.003$ & HR& \\
MB12323-OB120724\_M1 & $0.084 \pm 0.015$ & FS& \cite{MB120323} \\
OB060109 & $0.081 \pm 0.003$ & FS-PX& \cite{Bennett2010} \\
KB161820 & $0.080 \pm 0.009$ & FS-PX & \cite{KB161820}\\
MB140175-OB140676 & $0.079 \pm 0.075$ & PX & \cite{MB140175}\\
KB162142 & $0.074 \pm 0.009$ & FS-PX & \cite{KB161820}\\
MB130618-OB131721 & $0.064 \pm 0.016$ & FS & \cite{MB130618}\\
KB170016-OB171392-OB171434-MB170425 & $0.061 \pm 0.004$ & PX & \cite{KB170016}\\
MB11293 & $0.059 \pm 0.009$ & HR & \cite{MB11293}\\
KB150186-OB151670-OB151673-MB150379 & $0.058 \pm 0.015$ & FS & \cite{KB150186}\\
\hline
\end{tabular}
\end{table*}

The list of historic microlensing events also contains 51 published planetary events, which surely attract the greatest attention. In fact, the perspective of resolving the lens from the source for such events allows for a much more accurate mass estimate for the planet and the properties of the planetary system, including a more detailed view on the host star.

Known planetary events have already been investigated in detail and the results are publicly available in the respective papers. In particular, accurate models have been proposed taking into account follow-up data, re-reduced photometry and information from CMDs, which allow to exploit the measure of the finite-source effect and fix the angular Einstein angle. In some cases, with a good measure of both the finite-source effect and the microlensing parallax, it has been possible to derive the relative heliocentric proper motion between lens and source. Then, it has been sufficient to multiply by the time baseline to obtain predictions for the expected separation between lens and source. These values are reported in Table \ref{Tab planets}. When one of these effects is not measured, some Bayesian analysis is eventually needed to infer the properties of the microlensing event, including the proper motion. We have consequently flagged these events in our table. When more than one model fits the data, we have reported the corresponding expectations.

\begin{figure}
    \centering
    \includegraphics[width = 0.4\textwidth]{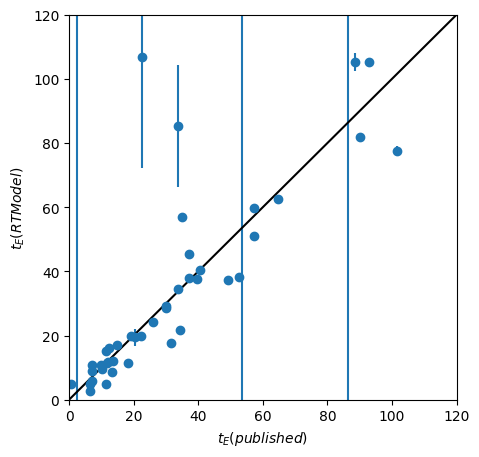}
    \caption{Comparison between the Einstein time obtained in our models and the value published in the discovery paper for the 51 planets in the EGBS.}
    \label{Fig tE planets}
\end{figure}

{ As can be seen from Fig. \ref{Fig tE planets}, all the expected separations reported in this table are in good agreement with the predictions made by our general re-modeling of the original survey data, which makes us more confident with respect to our modeling.} Consistently with our full sample of microlensing events, there are just 4 planetary events for which the lens should be separated by more than 0.1'', 18 events with a separation higher than 0.05'' and 30 events with separation higher than 0.025''.

The four planetary events for which a separation higher than one pixel is predicted are OGLE-2005-BLG-390 \citep{Beaulieu2006}, { MOA-2014-BLG-547/OGLE-2014-BLG-1760 \citep{MB140547}}, MOA-2013-BLG-260/OGLE-2013-BLG-0341 \citep{Gould2014} { and KMT-2016-BLG-0212 \citep{KB160212}}. The first is a twenty-year-old event, for which a high separation is quite expected. Unfortunately, the lens star should be a faint M-dwarf, while the source is a giant. A high contrast of about 10 magnitudes in this case makes the detection of the faint lens very challenging. { The second event has been recently resolved by Keck adaptive optics: in this case the \euclid image may be useful to confirm the lens detection. For the third event the lens system is relatively close and a high proper motion is derived in the models.} Since the source is relatively faint and the lens is made of a binary system of M-dwarfs, this event seems more promising for a successful detection. The fourth event in the list comes with four possible models in which the lens has mass between 0.36 and 0.45 $M_\odot$ and the companion could be a brown dwarf or a sub-Neptune depending on the chosen alternative. The different models predict different proper motions and different magnitudes for source and lens. Therefore, a resolution by \euclid should be able to distinguish these alternatives.

\section{Conclusions}

Decades of microlensing observations towards the bulge have accumulated tens of thousand microlensing events. Thanks to the \euclid Galactic Bulge Survey, we have the opportunity to resolve the lens from the source for a substantial fraction of all past microlensing events, bringing the contribution to the knowledge of the Galaxy from microlensing to a higher level.

In this investigation we have compiled the full list of historic events in the footprint of the EGBS. By re-modeling all events in a homogeneous way, we have revised the estimates for the basic microlensing parameters and we have issued predictions for the expected separation of lens and source for { 7801} microlensing events. In our catalog, astronomers can find all events with their coordinates, baseline magnitude, Einstein time and parallax (if measurable), along with an assessment of the possible category (single lens, binary lens or binary source). The reported chi square in the fit is also useful to discard events based on possible problems with systematics or poor modeling. For each event, we report the probability that the lens is separated from the source by more than one pixel, half pixel, a quarter pixel, or the size of the average PSF. It is clear that the individual cases may feature peculiarities that are not represented in our table and that may pose additional difficulties, such as high contrast between lens and source, a dark lens, additional blended stars, nearby saturated stars. All these can be evaluated and understood only when the real data are examined. However, our list represents the starting point and the reference for driving all future detailed analyses of individual microlensing events toward the most promising cases. 

We note there is currently an ongoing large {\it HST} survey of the Roman GBTDS fields (GO-17776; \citep{Terry2026b}). Much like the EGBS, a primary goal of this {\it HST} precursor survey is to extend the time baseline for which \roman lenses and sources can be studied with high-resolution imaging. Joint analysis of the EGBS and {\it HST} precursor data will allow for robust multi-passband characterization of the lenses in this historical event sample.

Finally, the \roman Galactic Time Domain Survey \citep{ROTAC2025} will be extremely useful to prolong the follow-up of historic events and confirm the conclusions of the analyses based on EGBS alone. At the present time, therefore, we are right in the middle of the path to the construction of a broad database of lens-source pairs that can be envisaged as an alternative way to obtain a 3D tomography along our line of sight to the Galactic bulge.

\begin{acknowledgements}
The OGLE project has received funding from the Polish National Science Centre grant OPUS-28 2024/55/B/ST9/00447 to AU.
The MOA project is supported by JSPS KAKENHI Grant Number JP16H06287, JP22H00153, JP23KK0060 and JP25H00668. 
This research has made use of publicly available data (https://kmtnet.kasi.re.kr/ulens/) from the KMTNet system operated by the Korea Astronomy and Space Science Institute (KASI) at three host sites of CTIO in Chile, SAAO in South Africa, and SSO in Australia. Data transfer from the host site to KASI was supported by the Korea Research Environment Open NETwork (KREONET). VB acknowledges financial support from PRIN2022 CUP D53D23002590006. This work was supported by the University of Tasmania through the endowed Warren Chair in Astronomy. AAC, J-PB, and ET NR have been supported by the Australian Government through the Australian Research Council Discovery Project Grants 200101909 and 240101842.  Research in France has been supported through the SPACE-MLENS ANR grant ANR-24-CE31-3263.
\end{acknowledgements}

%\section*{Data Availability}

\bibliographystyle{aa} 
\bibliography{Historic.bbl}
\label{LastPage}
\end{document}